\documentstyle[12pt,psfig]{article}
\catcode`\@=11
\def\marginnote#1{}
\newcount\hour
\newcount\minute
\newtoks\amorpm
\hour=\time\divide\hour by60
\minute=\time{\multiply\hour by60 \global\advance\minute by-
\hour}
\edef\standardtime{{\ifnum\hour<12 \global\amorpm={am}%
    \else\global\amorpm={pm}\advance\hour by-12 \fi
    \ifnum\hour=0 \hour=12 \fi
    \number\hour:\ifnum\minute<100\fi\number\minute\the\amorpm}}
\edef\militarytime{\number\hour:\ifnum\minute<100\fi\number\minute}
\def\draftlabel#1{{\@bsphack\if@filesw {\let\thepage\relax
  \xdef\@gtempa{\write\@auxout{\string
    \newlabel{#1}{{\@currentlabel}{\thepage}}}}}\@gtempa
    \if@nobreak \ifvmode\nobreak\fi\fi\fi\@esphack}
     \gdef\@eqnlabel{#1}}
\def\@eqnlabel{}
\def\@vacuum{}
\def\draftmarginnote#1{\marginpar{\raggedright\scriptsize\tt#1}}
\def\draft{\oddsidemargin -.5truein
        \def\@oddfoot{\sl preliminary draft \hfil
        \rm\thepage\hfil\sl\today\quad\militarytime}
        \let\@evenfoot\@oddfoot \overfullrule 3pt
        \let\label=\draftlabel
        \let\marginnote=\draftmarginnote

\def\@eqnnum{(\theequation)\rlap{\kern\marginparsep\tt\@eqnlabel}%
\global\let\@eqnlabel\@vacuum}  }
\def\preprint{\twocolumn\sloppy\flushbottom\parindent 1em
        \leftmargini 2em\leftmarginv .5em\leftmarginvi .5em
        \oddsidemargin -.5in    \evensidemargin -.5in
        \columnsep 15mm \footheight 0pt
        \textwidth 250mmin      \topmargin  -.4in
        \headheight 12pt \topskip .4in
        \textheight 175mm
        \footskip 0pt

\def\@oddhead{\thepage\hfil\addtocounter{page}{1}\thepage}
        \let\@evenhead\@oddhead \def\@oddfoot{} \def\@evenfoot{}
}
\def\titlepage{\@restonecolfalse\if@twocolumn\@restonecoltrue\onecolumn
     \else \newpage \fi \thispagestyle{empty}\c@page\z@
        \def\thefootnote{\fnsymbol{footnote}} }
\def\endtitlepage{\if@restonecol\twocolumn \else  \fi
        \def\thefootnote{\arabic{footnote}}
        \setcounter{footnote}{0}}  
\catcode`@=12
\relax
\def\be{\begin{equation}}
\def\ee{\end{equation}}
\def\bea{\begin{eqnarray}}
\def\eea{\end{eqnarray}}
\def\simlt{\stackrel{<}{{}_\sim}}
\def\simgt{\stackrel{>}{{}_\sim}}

\def\NPB#1#2#3{{\it Nucl.~Phys.} {\bf{B#1}} (19#2) #3}
\def\PLB#1#2#3{{\it Phys.~Lett.} {\bf{B#1}} (19#2) #3}
\def\PRD#1#2#3{{\it Phys.~Rev.} {\bf{D#1}} (19#2) #3}
\def\PRL#1#2#3{{\it Phys.~Rev.~Lett.} {\bf{#1}} (19#2) #3}
\def\ZPC#1#2#3{{\it Z.~Phys.} {\bf C#1} (19#2) #3}
\def\PTP#1#2#3{{\it Prog.~Theor.~Phys.} {\bf#1}  (19#2) #3}

\def\PR#1#2#3{{\it Phys.~Rep.} {\bf#1} (19#2) #3}
\def\RMP#1#2#3{{\it Rev.~Mod.~Phys.} {\bf#1} (19#2) #3}
\def\HPA#1#2#3{{\it Helv.~Phys.~Acta} {\bf#1} (19#2) #3}

\relax

\begin{document}
\topmargin-2.5cm
%
\begin{titlepage}
\begin{flushright}
CERN-TH/95-18\\
DESY 95-039 \\
IEM-FT-97/95 \\
hep--ph/9504241 \\
\end{flushright}
\vskip 0.3in
\begin{center}{\Large\bf
Improved metastability bounds on the \\ Standard Model Higgs mass
\footnote{Work supported in part by
the European Union (contract CHRX-CT92-0004) and
CICYT of Spain
(contract AEN94-0928).}  }
\vskip .5cm
{\bf J.R. Espinosa \footnote{Supported by
Alexander-von-Humboldt Stiftung.}
}\\
Deutsches Elektronen-Synchrotron DESY, Hamburg, Germany \\
and \\
\vskip.5cm
{\bf M. Quir\'os \footnote{On leave of absence from Instituto
de Estructura de la Materia, CSIC, Serrano 123, 28006-Madrid,
Spain.}}\\
CERN, TH Division, CH--1211 Geneva 23, Switzerland\\
\end{center}
\vskip.5cm
\begin{center}
{\bf Abstract}
\end{center}
\begin{quote}
Depending on the Higgs-boson and top-quark masses, $M_H$ and $M_t$,
the effective potential of the Standard Model at finite (and zero)
temperature can have a deep and unphysical stable minimum
$\langle \phi(T)\rangle$ at values of the field much larger than
$G_F^{-1/2}$. We have computed absolute lower bounds on $M_H$, as a
function of $M_t$, imposing the condition of no decay by thermal fluctuations,
or quantum tunnelling, to the stable minimum.
Our effective potential at zero temperature
includes all next-to-leading logarithmic corrections (making it
extremely scale-independent), and we have used pole masses for the
Higgs-boson and top-quark. Thermal corrections to the effective potential
include plasma effects by one-loop ring resummation of Debye masses. All
calculations, including the effective potential and the bubble nucleation
rate, are performed numerically, and so the results do not rely on any
kind of analytical approximation. Easy-to-use fits are provided for the
benefit of the reader. Conclusions on the possible Higgs detection at
LEP-200 are drawn.
\end{quote}
\vskip1.cm

\begin{flushleft}
CERN-TH/95-18\\
March 1995 \\
\end{flushleft}

\end{titlepage}
\setcounter{footnote}{0}
\setcounter{page}{0}
\newpage
%
\section{Introduction}

For a particular range of values of the
Higgs-boson and top-quark masses, $M_H$ and $M_t$,
the effective potential of the Standard Model (SM)
exhibits an unphysical stable minimum at values of the field much larger
than the electroweak scale. This effect is accentuated for large top
Yukawa coupling $h_t$, which drives the SM quartic coupling $\lambda$
to negative values at large scales. Therefore, the vacuum stability
requirement in the SM imposes a severe lower bound on $M_H$, which
depends on $M_t$ and the cutoff $\Lambda$ beyond which new physics
operates. This bound was computed in various approximations
[1--5], and,
more recently, using the improved one-loop effective potential including
all next-to-leading logarithm corrections and pole masses for the
Higgs-boson and the top-quark [6--8]. It was proved in ref.~\cite{CEQ}
that the latter effects can be very
specially important, in particular for large
top-quark masses (as the recent experimental evidence indicates
\cite{CDF,D0}) and
for low values of $\Lambda$ (which can be interesting for the future
range of masses that will be covered at LEP-200 \cite{KS}).

However, even if the lower bounds on $M_H$ arising from stability
requirements are a valuable indication, they cannot be considered as
absolute lower bounds in the SM since we cannot logically exclude the
possibility of the physical electroweak minimum being a metastable one,
provided the  probability, normalized with respect to the expansion rate
of the Universe, for decay to the  unphysical (true) minimum, be
negligibly small.
This we will call metastability requirement. A first
step in that direction was given in ref.~\cite{A}, where bounds on the
Higgs mass from the requirement of metastability of the electroweak vacuum
at finite temperature, for temperatures below the critical temperature
of the electroweak phase transition, were given. It was subsequently
noticed \cite{AV} that the strongest bounds come from the requirement of
metastability for temperatures higher than the electroweak critical
temperature. In that paper the effective potential was calculated in the
leading-logarithm approximation, with tree-level masses for
the Higgs-boson and top-quark, and using the high-temperature limit for
thermal corrections as well as semi-analytical approximations for the
calculation of the energy of the
critical bubble and so the tunnelling probability
by thermal fluctuations.

In view of the future Higgs search at LEP-200 and future colliders, it is
extremely important that the bounds provided
on the Higgs mass in the SM be as
accurate as possible. In this paper we will compute metastability lower
bounds on the Higgs mass as a function of the top mass and the cutoff
$\Lambda$ using:
\begin{itemize}
\item
an effective potential including next-to-leading logarithm corrections,
and guaranteeing to a large extent scale independence, as in \cite{CEQ};
\item
physical (pole) masses for the Higgs-boson and top-quark;
\item
thermal corrections to the effective potential including plasma effects
by one-loop resummation of Debye masses; these corrections are evaluated
numerically and thus do not rely on the high temperature expansion;
\item
numerical calculation of the bounce solution
and the energy of the critical bubble.
\end{itemize}

As a consequence of the previous input, our lower bound on $M_H$ reduces
dramatically with respect to the results of ref.~\cite{AV}. To fix the
ideas, for $M_t=175$ GeV, the bound reduces by $\sim$~10 GeV for
$\Lambda=10^4$ GeV, and $\sim$~30 GeV for $\Lambda=10^{19}$ GeV.

\section{The effective potential}

The starting point in our analysis is the effective potential of the
SM at finite temperature. It contains the usual zero-temperature term
and the thermal corrections \cite{Q} as:
\be
\label{pottotal}
V_{\rm eff}(\phi,T)=V_{\rm eff}(\phi,0)+\Delta V_{\rm eff}(\phi,T)\ .
\ee
The first term in (\ref{pottotal}) can be written in the 't Hooft--Landau
gauge and the $\overline{\rm MS}$ scheme as \cite{FJSE}:
\be
\label{sumain}
V_{\rm eff}(\phi,0)=\sum_{\rm L} V_{\rm L}(\phi)
\ee
where $V_{\rm L}$ is the L-loop correction to the effective potential at
zero-temperature, namely
\be
\label{vcero}
V_0=-\frac{1}{2}m^2(t)\phi^2(t)+\frac{1}{8}\lambda(t)\phi^4(t)
\ee
\be
\label{vuno}
V_1=\sum_{i=W,Z,t}\frac{n_i}{64\pi^2}m_i^4(\phi)
\left[\log\frac{m_i^2(\phi)}{\mu^2(t)}-C_i\right]+\Omega(t)\ ,
\ee
where
$$n_W=6,\ \ n_Z=3,\ \ n_t=-12,\ \
{\displaystyle C_W=C_Z=\frac{5}{6}},\ \
{\displaystyle C_t=\frac{3}{2}},$$
the masses are defined as usual by
$$m^2_W(\phi)=\frac{1}{4}g^2(t)\phi^2(t),\ \
m^2_Z(\phi)=\frac{1}{4}(g^2(t)+g'^2(t))\phi^2(t),\ \
m^2_t(\phi)=\frac{1}{2}h^2_t(t)\phi^2(t),$$
and $\Omega$ is the one-loop contribution
to the cosmological constant \cite{FJSE}, which will turn out to be
irrelevant in our calculation. Likewise the contribution of the Higgs and
Goldstone boson sector can be consistently included \cite{CEQR},
although it is numerically irrelevant.

In eqs.~(\ref{vcero}) and (\ref{vuno}) the parameters $\lambda(t)$ and $m(t)$
are
the SM quartic coupling and mass, whereas $g(t)$, $g'(t)$ and $h_t(t)$ are the
$SU(2)$, $U(1)$ and top Yukawa couplings respectively. All parameters are
running with the SM renormalization group equations (RGE). The Higgs field is
running as $\phi(t)=\xi(t)\phi$, with
$\xi(t)=\exp\{-\int_0^t\gamma(t')dt'\}$ where $\gamma(t)$ is the anomalous
dimension of the Higgs field. Finally the scale $\mu(t)$ is related to the
running parameter $t$ by $\mu(t)=\mu\exp(t)$, where $\mu$ is a given scale
that fixes the starting of the running and will be taken equal to the
physical $Z$ mass.

It has been shown \cite{JAP} that the L-loop effective potential improved
by (L+1)-loop RGE resums all L$_{\rm th}$-to-leading logarithm contributions.
Consequently we have considered \cite{CEQ} all the $\beta$- and
$\gamma$-functions of the previous parameters to two-loop order
\cite{FJSE} so that our calculation contains all next-to-leading
logarithmic corrections.

The potential (\ref{sumain}) has been proved \cite{CEQ,CEQR} to be
very scale-independent. It means that any judicious choice of the
scale $\mu(t)$ should give a very good and accurate numerical description
of it. In particular the choice $\mu(t)=\phi(t)$ minimizes the size of
radiative corrections and will be taken from here on.
The potential (\ref{sumain}) can
have, depending on the values of the Higgs-boson and top-quark masses,
other minima at large values of the field. In particular it was
shown that at any stationary point of (\ref{sumain}), $\phi_{\rm stat}=
2m^2/\tilde{\lambda}$, with
\bea
\label{lambdahat}
\tilde{\lambda}&=&\lambda-
{\displaystyle\frac{1}{16\pi^2}}\left\{6h_t^4\left[\log
{\displaystyle\frac{h_t^2}{2}}-1\right]-
{\displaystyle\frac{3}{4}}g^4\left[\log
{\displaystyle\frac{g^2}{4}}-{\displaystyle\frac{1}{3}}
\right]\right.\nonumber \\
&&\nonumber \\
&-&\left.{\displaystyle\frac{3}{8}}\left(g^2+g'^2\right)^2
\left[\log{\displaystyle\frac{
\left(g^2+g'^2\right)}{4}}-{\displaystyle\frac{1}{3}}\right]
\right\}.
\eea
For stationary points much larger than the electroweak scale,
$\tilde{\lambda}\ll 1$ and the curvature of the potential is given by
\be
\label{curvatura}
\left.\frac{\partial^2 V_{\rm eff}}{\partial \phi^2(t)}\right|_{\phi(t)=
\phi_{\rm stat}(t)}=\frac{1}{2}\left(\beta_{\lambda}
-4\gamma\lambda\right)\phi^2_{\rm stat}\simeq \frac{1}{2}\beta_{\lambda}
\phi^2_{\rm stat}\ ,
\ee
where all the parameters are evaluated at the scale $\phi_{\rm stat}$.
Eq.~(\ref{curvatura}) shows that the stationary point is a minimum
(maximum) if $\beta_{\lambda}>0$ ($\beta_{\lambda}<0$).

In fact, the large field structure
of the effective potential at zero-temperature
can be understood from (\ref{curvatura}). Typically, for low scales
$\beta_{\lambda}$ is negative and dominated by the term $-12h_t^4$. This
means that $\tilde{\lambda}$ is decreasing as the scale increases and
will satisfy the condition $\tilde{\lambda}\sim 0$ for a stationary point,
which will turn out to be a maximum since $\beta_{\lambda}$ is negative.
However, as the top coupling $h_t$ decreases with the scale,
and the gauge couplings provide a positive contribution to $\beta_{\lambda}$,
at a given scale the contribution of top and gauge couplings will balance
and there will be a turn over at a particular scale from negative
to positive $\beta_{\lambda}$. This will make $\tilde{\lambda}$ to
increase and cross zero, indicating the presence of a minimum, since
now $\beta_{\lambda}>0$.

The previous pattern translates into a well-defined
structure of the effective potential at zero-temperature. The locations
of maxima and minima depend on the SM parameters, in particular on the
Higgs-boson and top masses. Stability bounds can be established on the
basis that the maximum \footnote{Strictly speaking we should replace
in this sentence "the maximum of the potential" by
the "value of the field where the potential
is deeper than the electroweak minimum". However the difference is
numerically irrelevant given the sharp descent of the potential after the
maximum.} occurs for values of the field larger than the cutoff $\Lambda$
beyond which the SM is no longer valid.

Using pole (physical) values of the Higgs mass $M_H$:
\be
\label{MHphys}
M_H^2= m_{H}^2(t) +{\rm Re}[\Pi (p^2=M_H^2)-\Pi (p^2=0)],
\ee
where $m_H(t)$ is the running mass defined as the second derivative of
the effective potential and
 $\Pi(p^2)$ is the renormalized self-energy of the Higgs-boson
\footnote{The one-loop $t$-dependence of (\ref{MHphys}) drops out.
Explicit expressions for
$m_H^2(t)$ and $\Pi(M_H^2)-\Pi(0)$ can be found in ref.~\cite{CEQR}.},
and top-quark mass \cite{gray} $M_t$:
\be
\label{mtphys}
M_t=\left\{1+{\displaystyle\frac{4}{3}}
{\displaystyle\frac{\alpha_S(M_t)}{\pi}}+
\left[16.11-1.04\sum_{i=1}^{5}\left(1-\frac{M_i}{M_t}\right)\right]
\left(\frac{\alpha_S(M_t)}{\pi}\right)^2
\right\} m_t(M_t),
\ee
where $M_i$, $i=1,\ldots,5$, represent the masses of the five lighter quarks,
lower bounds on $M_H$ as a function of $M_t$ and $\Lambda$
were put in \cite{CEQ}. We will
later on compare these bounds with those that will be obtained in this paper.
In Fig.~1 we show the shape of the effective
potential at zero-temperature (thick solid line) for $M_t=175$ GeV and
$M_H\sim 122$ GeV. We see that there is a deep minimum at large values of
$\phi$. Requiring absolute stability of the effective potential leads
to a rejection of these values of $M_t$ and $M_H$, as can also be
seen in \cite{CEQ}.

The thermal correction to Eq.~(\ref{pottotal}) can be computed using the
rules of field theory at finite temperature \cite{DJW,K}. Including
plasma effects \cite{Q,DJW,K,GPY}
by one-loop ring resummation of Debye masse \footnote{This approximation
is good enough for our purposes in this paper, since the infrared problem
for $\phi=0$ will not affect the phase transition to values of the
field $\gg G_F^{-1/2}$.}, it can be written as
\be
\label{vT}
\Delta V_{\rm eff}(\phi,T)=V_1(\phi,T)+V_{\rm ring}(\phi,T)\ .
\ee
The first term in (\ref{vT}) is the one-loop thermal correction
\be
\label{v1T}
V_1(\phi,T)=\frac{T^4}{2\pi^2} \left[
\sum_{i=W,Z} n_iJ_B\left(\frac{m_i^2(\phi)}{T^2}\right)
+n_tJ_F\left(\frac{m_t^2(\phi)}{T^2}\right)
\right]
\ee
where the functions $J_B$ and $J_F$ are defined by
\be
\label{jb}
J_B(y)=\int_0^{\infty} dx\ x^2\log\left[1-e^{-\sqrt{x^2+y^2}}\right]
\ee
and
\be
\label{jf}
J_F(y)=\int_0^{\infty} dx\ x^2\log\left[1+e^{-\sqrt{x^2+y^2}}\right]\ .
\ee

Plasma effects in the leading approximation can be accounted for by the
one-loop effective potential improved by the daisy diagrams
\cite{DJW,GPY}. This approximation takes into account the contribution
of hard thermal loops in the higher-loop expansion. The second term of
Eq.~(\ref{vT}) is given by \cite{ring}
\be
\label{ring}
V_{\rm ring}(\phi,T)=\sum_{i=W_L, Z_L, \gamma_L}n_i
\left\{\frac{m_i^3(\phi)T}{12\pi}
-\frac{{\cal M}_i^3(\phi)T}{12\pi} \right\}\ ,
\ee
where only the longitudinal degrees of freedom of gauge bosons,
$$\frac{1}{2}n_{W_L}=n_{Z_L}=n_{\gamma_L}=1, $$
are accounted. The Debye-corrected masses are given by
\bea
\label{debye}
{\cal M}^2_{W_L}& = & m^2_W(\phi)+\frac{11}{6}g^2T^2 \nonumber \\
{\cal M}^2_{Z_L}& = & \frac{1}{2} \left[m_Z^2(\phi)+\frac{11}{6}
\frac{g^2}{\cos^2\theta_W}T^2+\Delta(\phi,T) \right] \\
{\cal M}^2_{\gamma_L}& = & \frac{1}{2} \left[m_Z^2(\phi)-\frac{11}{6}
\frac{g^2}{\cos^2\theta_W}T^2+\Delta(\phi,T) \right] \nonumber
\eea
where the discriminant
\be
\label{delta}
\Delta^2(\phi,T)=m_Z^4(\phi)+\frac{11}{3} \frac{\cos^22\theta_W}{\cos^2
\theta_W}\left[m_Z^2(\phi)+\frac{11}{12}\frac{g^2}{\cos^2\theta_W}T^2
\right]T^2
\ee
is responsible for the rotation at finite temperature from the basis
$(W_{3L},B_L)$ to the mass eigenstate basis $(Z_L,\gamma_L)$.

In Fig.~1 we plot the effective potential at
$T=T_t=2.5\times10^{15}$ GeV
(thin solid line), for the previously considered values of
$M_t$ and $M_H$, where the temperature $T_t$
will be defined in the next sections.
We can see that for values of the
field such that $\phi\simlt 10 T_t\sim 10^{16}$ GeV
the thermal corrections dominate over the zero-temperature
term, while for $\phi\simgt 10^{16}$ GeV, thermal corrections
are exponentially suppressed and the potential drops to the
zero-temperature value. For $T\gg T_t$ the minimum disappears
and symmetry is restored.

\section{The thermal tunnelling}

In a first-order phase transition, such as that depicted in Fig.~1,
the tunnelling probability rate per unit time per unit volume is given
by \cite{Langer,Linde}
\be
\label{rateT}
\frac{\Gamma}{\nu}\sim \omega T^4 e^{-E_b/T}\ ,
\ee
where, for our purposes the prefactor $\omega$ can be taken to be ${\cal
O}(1)$ as will be explained later, and $E_b$ (the energy of a bubble of
critical size) is given by the three-dimensional euclidean action $S_3$
evaluated at the {\it bounce} solution $\phi_B$
\be
\label{exponenteT}
E_b=S_3[\phi_B(r)].
\ee

At very high temperature the bounce solution has $O(3)$ symmetry, and
the euclidean action is provided by
\be
\label{S3}
S_3=4\pi \int_0^{\infty} r^2 dr\left[\frac{1}{2}\left(\frac{d\phi}
{dr}\right)^2+V_{\rm eff}(\phi(r),T) \right]\ ,
\ee
where $r^2=\vec{x}^2$,  the potential is normalized as $V_{\rm eff}(0,T)=0$,
and the bounce $\phi_B$ satisfies the Euclidean
equation of motion
\be
\label{eqmotT}
\frac{d^2 \phi}{dr^2}+\frac{2}{r}\frac{d\phi}{dr}=
\frac{dV_{\rm eff}(\phi,T)}{d\phi}
\ee
with the boundary conditions \footnote{For $T<T_c^{\rm EW}$ the first condition
in
(\ref{boundaryT}) should be replaced by  $\phi(\infty)=v^{\rm EW}(T)$.}
\bea
\label{boundaryT}
\lim_{r\rightarrow\infty}\phi(r)& = & 0 \nonumber \\
\left.\frac{d\phi}{dr}\right|_{r=0} & = & 0 .
\eea

The semiclassical picture is that unstable bubbles
(either expanding or collapsing) are nucleated behind the barrier,
at $\phi_B(0)$, with a probability rate given by (\ref{rateT}).
Whether or not they fill the Universe depends on the relation between
the probability rate (\ref{rateT}) and the expansion rate of the
Universe. The actual probability $P$ is obtained by multiplying the
probability rate (\ref{rateT}) by the volume of our current horizon
scaled back to the temperature $T$ and by the time the Universe spent at
temperature $T$ \cite{A}. One then obtains
\be
\label{probabilidad}
\frac{dP}{d\log T}=\kappa \frac{M_{P\ell}}{T} e^{-E_b/T}
\ee
where
\be
\label{numero}
\kappa\sim 3.25\times 10^{86}\ .
\ee
The total integrated probability is defined as
\be
\label{integral}
P(T_c)=\int^{T_c}_0 \frac{dP(T')}{dT'}\ dT',
\ee
where $T_c$ is the temperature at which the two minima of the
effective potential become degenerate. In fact, when $T\rightarrow T_c$
the probability rate goes to zero, since $E_b(T)\rightarrow\infty$.

Let us notice that the  total probability $P\equiv P(T_c)$
is not normalized to unity. In fact
the physical meaning of the integrated probability was discussed in
ref.~\cite{GW}, where it was shown that the fraction of space in the
(old) metastable phase in a first-order phase transition is given by
\be
\label{old}
f_{\rm old}=e^{-P}
\ee
and so the fraction of space in the (new) stable phase is
\be
\label{new}
f_{\rm new}=1-e^{-P}\ .
\ee
In this way for values of $P\ll 1$
all the space is in the metastable phase,
while for $P\gg 1$ all the space
is in the stable phase. We will see that
the critical value of the probability, $P={\cal O}(1)$,
can simply be taken
as the condition for the space to be
in the metastable phase since, as a function of $M_H$, $P$ is very
rapidly (exponentially) varying.

In Figs.~2 and 3 we show plots of $E_b$ and $dP/d\log_{10}T$,
respectively, versus $\log_{10}T$ for $M_t=175$ GeV and
$M_H\sim 122$ GeV, as in Fig.~1. We see that there is a minimum of $E_b/T$ for
a
temperature of $\log_{10}(T/{\rm GeV})\sim 17.5$
and, correspondingly, a maximum of
$dP/d\log_{10}T$ for $\log_{10}(T/{\rm GeV})\sim 16.5$. The fact that the two
stationary points do not coincide is a consequence of the prefactor in
(\ref{probabilidad}), and proves that integrating (\ref{probabilidad}) by
the steepest-descent method around the point
$B'(T)=0$, $B(T)\equiv E_b(T)/T$, as can
usually be found in the literature \cite{GW}, is not a good approximation
in our case. We have therefore performed the integral of (\ref{probabilidad})
numerically. In Fig.~4
we have plotted the detail of the effective potential for the same values
of $M_t$ and $M_H$ and the temperature $T_t$. We have shown, with
the arrow on the tip of the wavy line, the location of the bounce solution
$\phi_B(0)\sim 1.9\times 10^{16}$ GeV at this temperature.

As we can see from Fig.~3, the total integrated probability
for the considered values of $M_t$ and $M_H$
is greater than 1.
However, it may also happen to be smaller than 1. This will be the case for
any fixed value of $M_t$ and sufficiently large values of $M_H$. For those
values of $M_t$ and $M_H$,
even if there is a metastable minimum at the origin, there is no
dangerous transition to the deep stable minimum:
in spite of appearances,
the SM is robust for those values of $M_t$ and $M_H$. If, for the same
value of $M_t$, we decrease the value of $M_H$, then the total integrated
probability increases, until it reaches values of ${\cal O}(1)$
in which case $f_{\rm new}\sim 1$ and the phase transition takes
place.

The example that have been worked out in Figs.~1--4 have been tuned to get
the bound on $M_H$ for $M_t=175$ GeV. For $M_H \simlt 122$ GeV the integrated
probability increases and the phase transition always
takes place. But, how
sensitive is the obtained bound on $M_H$ to the precise definition of the
(critical) probability for the onset of the phase transition? The answer is,
very little. This is illustrated in Fig.~5 where we plot $\log P$, as a
function of $M_H$ for $M_t=175$ GeV. We can see that $\log P$
crosses zero
at $M_H\sim 122$ GeV, while $\Delta M_H\sim\pm 1$ GeV corresponds to
$\Delta\log P\sim\pm 20$, which means that the error induced by the precise
definition of the critical probability is negligible. This also means
that no precise knowledge of the prefactor $\omega$ is
required to get an accurate value of the bound on $M_H$. In fact, it has
been shown \cite{omega} (computing the Higgs fluctuations on the
background of the critical bubble) that for the electroweak phase transition
the
prefactor $\omega$ might significantly suppress the
tunnelling rate. In our case the quartic Higgs coupling in the relevant
range of scales is very small and these fluctuations will have no such a
dramatic effect \footnote{Even ignoring this fact,
the results of \cite{omega} give $\log\omega\sim -23$ (when
the height of the barrier is small compared with the free-energy difference
between the minima, as in our case), and Fig. 5 shows that this effect
would change the bound on $M_H$ by less than $1\ GeV$.}.

\section{The bounds}

In the previous section we have worked out in detail the bound
on the Higgs mass ($M_H\simgt 122$ GeV) for a particular value of the
top-quark mass ($M_t=175$ GeV) assuming  implicitly a SM cutoff at
a scale equal to $10^{19}$ GeV (see Fig.~5).  However the bounds should also
depend on the actual value assumed for $\Lambda$.

We start from values of $M_t$ and $M_H$ such that $P=1$
in (\ref{integral}) for $\Lambda=10^{19}$ GeV. [This is, e.g., the case
illustrated in Fig.~5 for $M_t=175$ GeV and $M_H\sim 122$ GeV.]
Keeping  now $M_t$ fixed,  $P$ increases when
$M_H$ decreases. [See again Fig.~5 for illustration.]
This means that for those values of
$M_H$, and  $\Lambda=10^{19}$ GeV, the phase transition should take place,
as expected.
However, for any different fixed value of $\Lambda$, $\Lambda<10^{19}$ GeV,
we have to cutoff the integral
(\ref{integral}) such that $\phi_B(0)$ at the maximal integration temperature
$T_t$ satisfies \footnote{In fact Eq.~(\ref{condic}) is our definition of
$T_t$.}
\be
\label{condic}
\phi_B(0)\sim\Lambda
\ee
The phase transition will take place for those values of $M_H$ such that
\be
\label{cond}
P(T_t)\simgt 1\ .
\ee
In particular saturation of condition (\ref{cond})
leads to the actual bound on $M_H$ for $M_t$ and $\Lambda$
fixed.

{}From Figs.~1 and 4, and the results described
in the previous section, we deduce that the obtained bound of $M_H\sim 122$ GeV
for $M_t=175$ GeV corresponds to a maximal temperature of
$T_t=2.5\times 10^{15}$ GeV (see Fig.~3) and a corresponding
bounce of $\phi_B(0)\sim 10^{16}$ GeV and
a cutoff scale of $\Lambda\sim 10^{16}$ GeV.
Moving down with $\Lambda$, and keeping $M_t$ fixed, one should move down
with $M_H$ to saturate condition (\ref{condic}). This behaviour is
illustrated in Fig.~6, where we plot the lower bound on $M_H$ as a
function of $\Lambda$ for fixed values of $M_t$  from 140~GeV to
200~GeV, and $\alpha_S(M_Z)$ = 0.124. The upper curve corresponds
to $M_t=200$ GeV and the lower curve to $M_t=140$ GeV.
{}From Fig.~6 we can see that a measurement of $M_H$ and $M_t$ could give
under certain circumstances, an {\it upper} bound on the scale of new
physics. Given the present LEP bound on $M_H$ \cite{MH},
$M_H>64.3$ GeV (95\% c.l.), only if
$M_t>150$ GeV we could obtain an upper bound on the scale of new physics
from the Higgs detection and mass measurement. For instance if we fix
$M_t=200$ GeV we will obtain an upper bound on $\Lambda$ provided that
$M_H<175$ GeV. We will comment briefly,
in the next section, on the implications of this fact for
the Higgs search at LEP-200.

In Fig.~7 we have plotted the lower bound on $M_H$ as a function of
$M_t$ for different values of $\Lambda$: from
$\log_{10}[\Lambda/{\rm GeV}]$~=~4 (lower
solid) to 19 (upper solid). We
also present in Fig.~7, for the sake of comparison, the lower bounds arising
from the absolute stability requirements \cite{CEQ} for
$\log_{10}[\Lambda/{\rm GeV}]$~=~3
(lower dashed), 4 and 19 (upper dashed). We can
see that the solid curve corresponding to
$\log_{10}[\Lambda/{\rm GeV}]$~=~3 has
disappeared, which means that for this value of $\Lambda$, even if the
electroweak minimum can be metastable (this corresponds to the
region below the lower
dashed line), it never decays into the stable unphysical minimum.
The modification that the metastability bounds impose on the picture
where only absolute stability was imposed can be easily traced back from
Fig.~7. For $\Lambda=10^{19}$ GeV it is negligible for $M_t=200$ GeV,
while it can be as large as 25 GeV for $M_t=140$ GeV. However, for small
values of $\Lambda$ the modification is dramatic for the
interesting  range of
$M_t$ considered. For instance, for $\Lambda=10^4$ GeV the difference is
$\sim$~30
GeV for $M_t=200$ GeV and $\sim$~50 GeV for $M_t=165$ GeV.

We have made a linear fit to the solid curves of Fig.~7 as:
\be
\label{fit1}
M_H/{\rm GeV}=A(\Lambda)(M_t/{\rm GeV})-B(\Lambda)\ ,
\ee
where the coefficients $A$ and $B$ are given in the table. We have taken
$\alpha_S(M_Z)=0.124$ and the fit is accurate to 1 GeV, for $M_H>60$ GeV,
i.e. 150 GeV$< M_t <$~200 GeV.

\vspace{.5cm}
\begin{center}
\begin{tabular}{||c|c|c||}
\hline
$\log_{10}(\Lambda/{\rm GeV})$ & $A(\Lambda)$ & $B(\Lambda)$ \\
\hline
4 & 1.219 & 157 \\
5 & 1.533 & 186 \\
7 & 1.805 & 212 \\
9 & 1.958 & 230 \\
11 & 2.071 & 245 \\
13 & 2.155 & 258 \\
15 & 2.221 & 268 \\
19 & 2.278 & 277 \\
\hline
\end{tabular}
\end{center}
\vspace{0.5cm}
\begin{center}
Table: Coefficients $A(\Lambda)$ and $B(\Lambda)$ of Eq.~(\ref{fit1}).
\end{center}

\vspace{.5cm}
The dependence on $\alpha_S(M_Z)$ is illustrated in Fig.~8, where we
have taken \cite{alfa3}
\be
\label{alpha}
\alpha_S(M_Z)=0.124 \pm 0.006\ ,
\ee
fixed $\Lambda$ to its maximum physically interesting value of $10^{19}$ GeV,
and represented
the lower bound on $M_H$ for the central value of $\alpha_S$ in (\ref{alpha})
(diagonal thick solid line) and the two extreme values (diagonal thick
dashed lines). A fit to these lines, accurate to 1 GeV for $M_H>60$~GeV,
as those in (\ref{fit1}), is given by,
\be
\label{fit2}
M_H/{\rm GeV}=\left[2.278-4.654\left(\alpha_S-0.124\right)\right]
\left(M_t/{\rm GeV}\right)-277\ .
\ee

We have also shown in the plot the bounds corresponding to the requirement
of absolute stability \cite{CEQ} (diagonal thin lines) and, for the sake of
comparison,
the absolute upper bound in the MSSM \cite{CEQR} corresponding to the same
values of
$\alpha_S$ and to $\Lambda_{\rm SUSY}=1$~TeV.

\section{Conclusions}

We have obtained absolute lower bounds on the SM Higgs mass
$M_H$ as a function of the top-quark mass $M_t$, and the scale
$\Lambda$ beyond which the SM is no longer valid, from the
requirement of no decay by thermal fluctuations from the
metastable minimum at the origin to the true (deep) minimum
at large values of the field.
The bounds from the similar requirement of no
tunnelling by quantum fluctuations from the electroweak minimum
at zero-temperature are always weaker than the former ones.
For completeness we present them in Fig.~9, which should be
compared with the solid lines in Fig.~7.

Now we will
comment on the accuracy of our results. There are two types of
uncertainties: theoretical and experimental. We find that the former
are negligible as compared to the latter.
The theoretical uncertainties, leaving apart the very precise treatment of the
numerical analysis, include the definition of the critical probability, and
the possible gauge dependence of the result. As for the former, we have
seen in Fig.~5 that the result is completely insensitive to the
precise definition of the critical probability. Any value of ${\cal O}(1)$
would give the same result
\footnote{In fact the uncertainty in the determination of the Higgs mass
from the effective potential at zero temperature, even if bounded by
$\simlt 1$ GeV in our treatment \cite{CEQ}, is much greater than the
uncertainty from the definition of the critical probability. }.
A related uncertainty comes from the pre-factor $\omega$ in the
probability rate (\ref{rateT}). We expect this uncertainty to be comparable in
size to that associated with the definition of the critical probability,
and thus negligible when translated into an error in the determination of
the Higgs mass.
As for the gauge dependence, we expect it to affect our results
very little. In fact, as can be seen from
Figs.~1 and 4, the total effective potential at finite temperature is
totally dominated by the thermal correction, for $\phi<\phi_B(0)$. The
thermal correction (\ref{v1T}), coming from gauge bosons and
the top-quark, as
well as the Debye masses (\ref{debye}), are gauge-independent \cite{K},
while all the gauge dependence is encoded in the one-loop contribution
from the Higgs and Goldstone bosons to (\ref{v1T}), which we have
neglected, since it is numerically irrelevant \footnote{Neglecting the
scalar sector in radiative corrections is a normal
procedure for analysing the electroweak phase transition \cite{AH}. In
our case this approximation is especially justified since $\lambda\ll 1$
in the region near the stable (unphysical) minimum.}. Finally we should
mention that the very definition of the cutoff, or new physics
scale, $\Lambda$ has, itself, a fudge factor which increases when the
value of $\Lambda$ decreases. This is associated with the existence of
threshold effects at the scale $\Lambda$, necessary to match the SM
below $\Lambda$ with the new physics beyond $\Lambda$.  In practice
this effect should affect negligibly our results for high values of the
scale $\Lambda$.

The experimental uncertainties come from the uncertainty in the determination
of $\alpha_S(M_Z)$, which we will take as in (\ref{alpha}), and the
uncertainty in the (future) measurement of $M_t$. Normalizing the latter
as $M_t=176\pm 13$ GeV \cite{CDF}, we can write the uncertainty in the
presented lower bound for $\Lambda=10^{19}$ GeV, as

\be
\label{uncer}
{\displaystyle \Delta M_H=4.91\left(\frac{\Delta\alpha_S}{0.006}\right)
\left(\frac{M_t}{176}\right)+29.6\left(\frac{\Delta M_t}{13}\right) }\ ,
\ee
where all masses are expressed in GeV. Let us notice that the
uncertainty in (\ref{uncer}) gets reduced for $\Lambda < 10^{19}$ GeV.
In particular for $\Lambda=10^4$ GeV the factor 4.91 in (\ref{uncer})
becomes $\sim$~1 and the factor 29.6 becomes $\sim$~17.
A quick glance at (\ref{uncer}) shows that only a very precise determination
of $\alpha_S$ and $M_t$ can help in reducing the uncertainty on the
bound of $M_H$.

On the other hand, one can easily extract
information on the scale of new physics $\Lambda$ from a possible
measurement of the Higgs mass at LEP-200 and an experimental lower
bound on $M_t$. In fact, from Eq.~(\ref{fit2}) and the shape of bounds in
Fig.~6 as functions of $\Lambda$ we can deduce that the measurement of
$M_H$ would translate into an upper bound on the scale of
new physics provided that $M_t$ satisfies
\footnote{Notice that we have used in (\ref{fit3}) the 1$\sigma$ bound
on $\alpha_S(M_Z)$ in (\ref{alpha}), i.e. $\alpha_S<0.130$.}
\be
\label{fit3}
M_t>\frac{M_H}{2.25} + 123 \ ,
\ee
where all masses are in GeV. That is, from (\ref{fit3}) and the present
bound on $M_H$, we obtain that $M_t>152$ GeV is necessary
for a future Higgs-mass measurement to imply an upper bound
on the scale of new physics.
Non-detection of the Higgs at LEP-200, i.e. $M_H>90$ GeV, would imply
similarly $M_t>163$ GeV as a necessary condition to obtain an upper bound
on the scale of new physics from a future Higgs-mass measurement.
On the other hand, assuming experimental values for
the Higgs and top-quark masses,
$M^{\rm exp}_t\pm\Delta M^{\rm exp}_t$ and
$M^{\rm exp}_H\pm\Delta M^{\rm exp}_H$, the condition for an upper bound
$\Lambda_{\rm max}$ on
new physics is provided by Eq.~(\ref{fit3}), with
\be
\label{uno}
M_t=M^{\rm exp}_t-\Delta M^{\rm exp}_t
\ee
and
\be
\label{dos}
M_H=M^{\rm exp}_H+\Delta M^{\rm exp}_H
\ee
and the precise value of $\Lambda_{\rm max}$ is given by the intersection
of (\ref{uno}) and (\ref{dos}) in Fig.~6.

Finally, notice that the Higgs mass measurement might serve
(depending on the top mass) to disentangle between the SM,
with a cutoff at $\Lambda=10^{19}$ GeV, and the MSSM
with $\Lambda_{\rm SUSY}\gg M_Z$ (in which case the couplings of the
lightest Higgs are indistinguishable \cite{HH}
from the couplings of the SM Higgs). In Fig.~8 we have plotted the
upper bound on the light-Higgs mass in the MSSM for
$\Lambda_{\rm SUSY}\simlt 1$ TeV. We can see that only if
$M_t\simgt 180$ GeV there is a mass gap between the MSSM and the SM,
and measurement of $M_H$ will always exclude at least one of these
models. However, for $M_t\simlt 180$ GeV there is a large overlapping region
where both models would be indistinguishable.

\section*{Acknowledgements}
We thank W. Buchm\"uller and
M. Shaposhnikov for useful comments and discussions, and
J.~Ignatius and J.M.~Moreno for their help in handling critical bubbles.

\begin{figure}
\centerline{
\psfig{figure=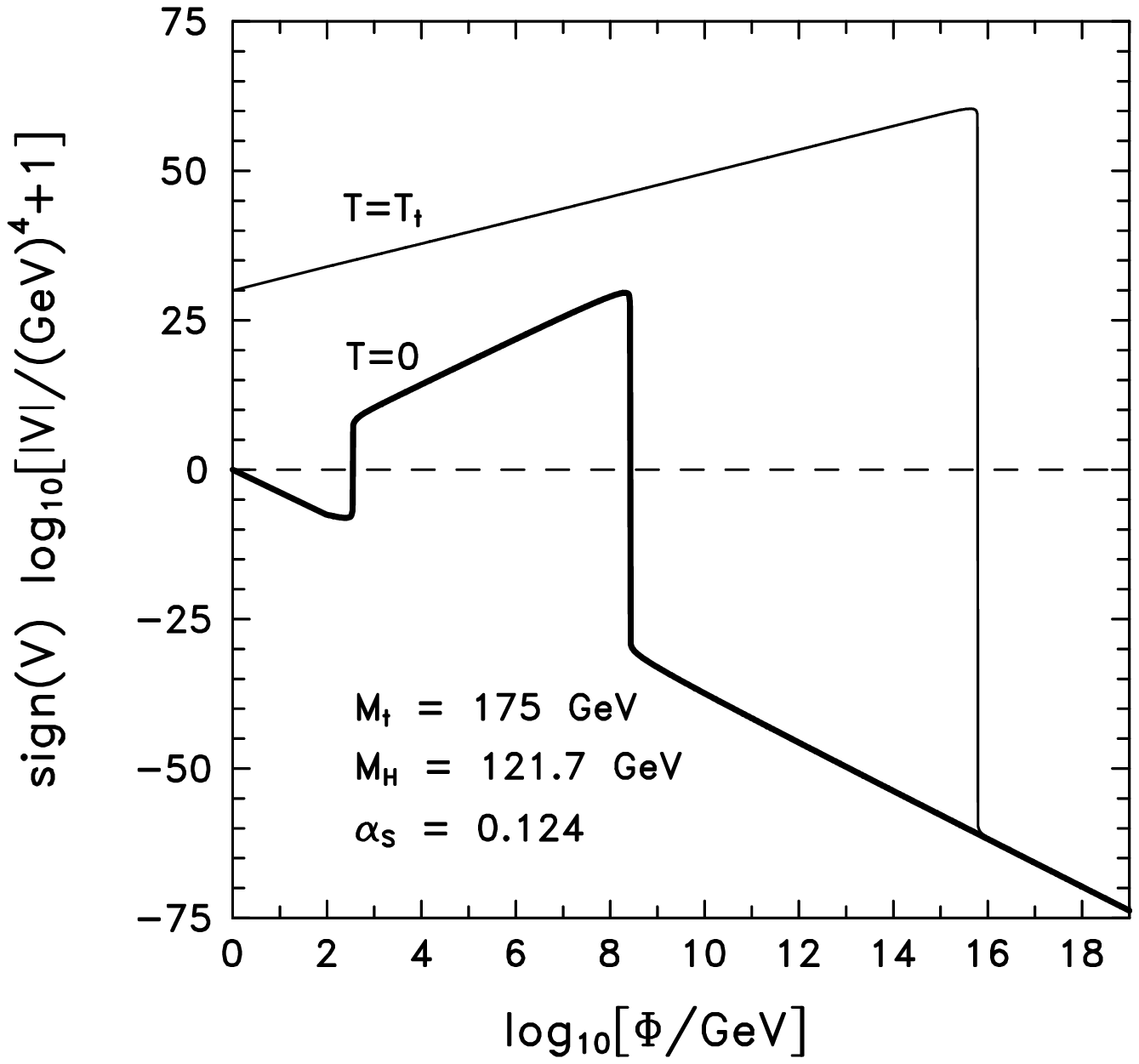,height=13cm,bbllx=6.cm,bblly=.cm,bburx=15.5cm,bbury=13cm}}
\caption{Plot of the effective potential for $M_t=175$ GeV, $M_H\sim 122$
GeV at $T=0$ (thick solid line) and $T=T_t=2.5\times 10^{15}$ GeV (thin solid
line).}
\end{figure}
\begin{figure}
\centerline{
\psfig{figure=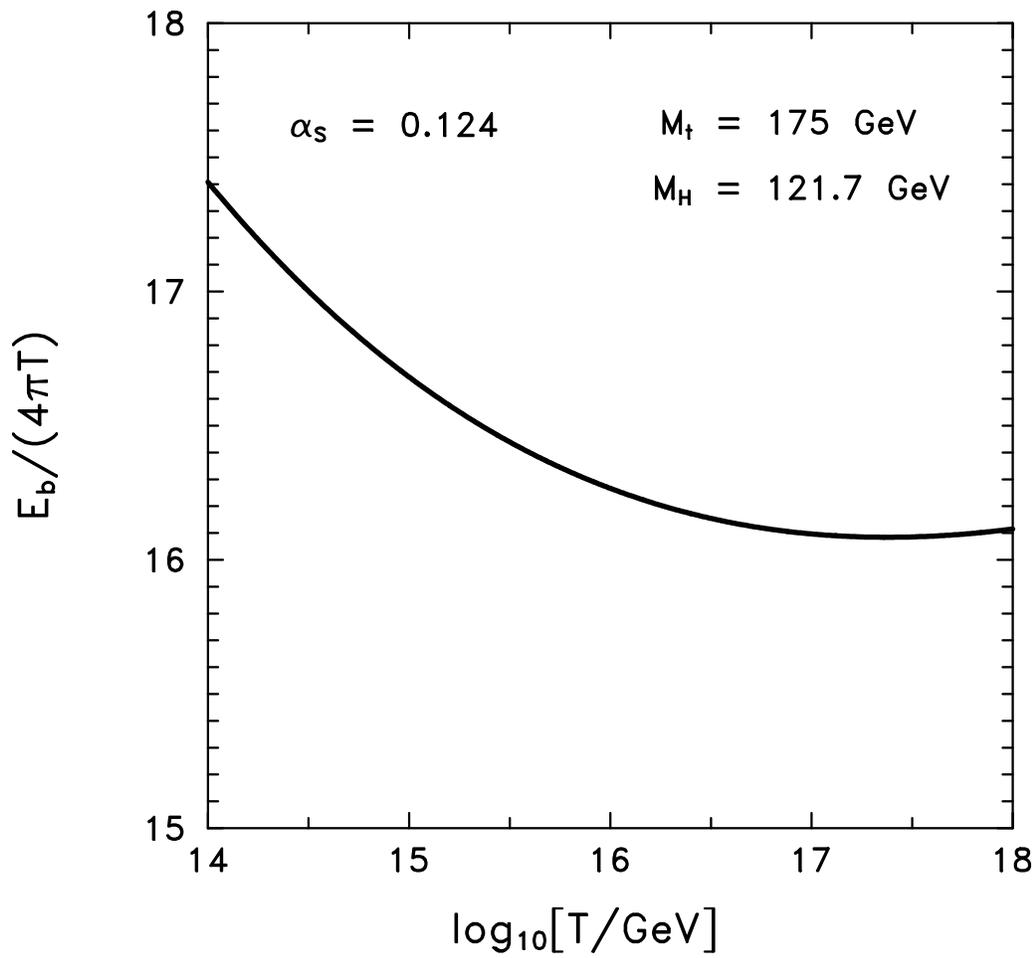,height=13cm,bbllx=6.cm,bblly=.cm,bburx=15.5cm,bbury=13cm}}
\caption{Plot of $E_b$, the energy of the critical bubble, as a function
of the temperature for the same values of $M_t$ and $M_H$ as in Fig.~1.}
\end{figure}
\begin{figure}
\centerline{
\psfig{figure=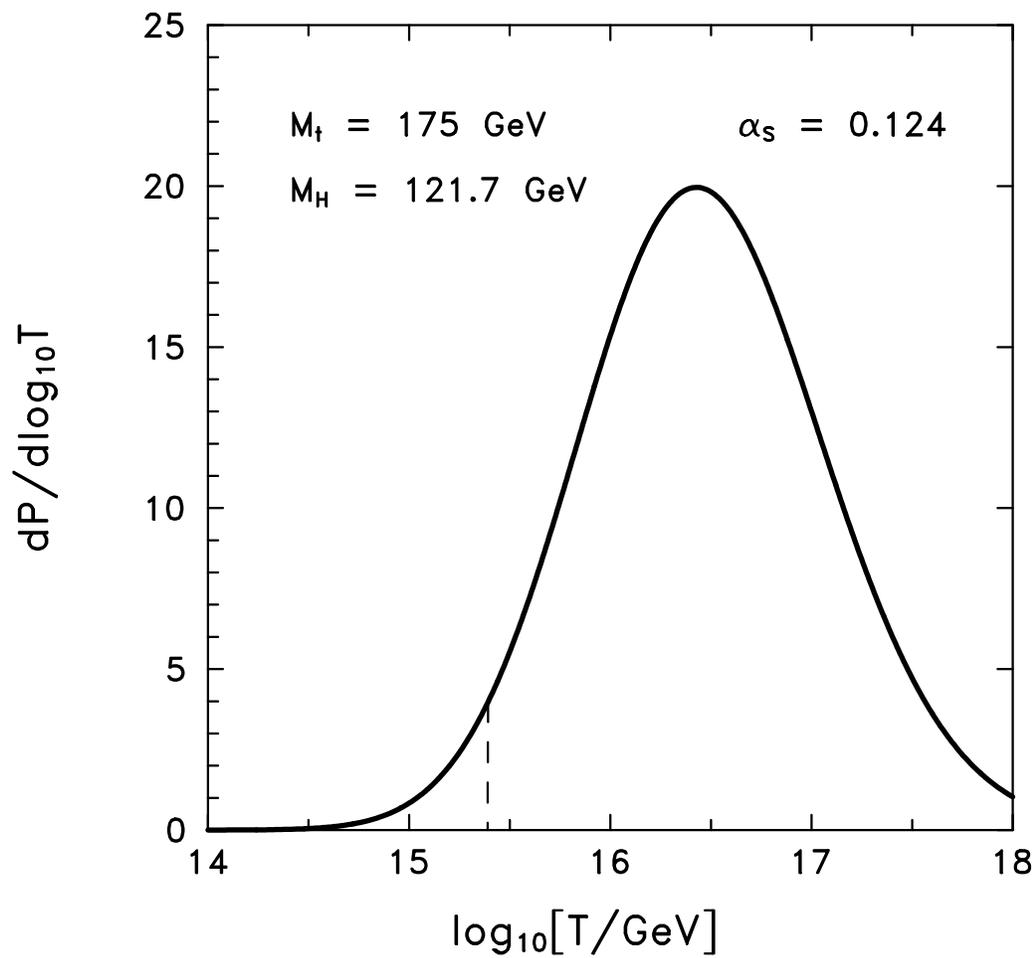,height=13cm,bbllx=6.cm,bblly=.cm,bburx=15.5cm,bbury=13cm}}
\caption{Plot of $dP/d\log_{10}T$ as a function of the temperature
for the same values of $M_t$ and $M_H$ as in Fig.~1. The temperature
$T_t=2.5\times 10^{15}$ GeV at which the integrated probability is equal to 1
is indicated with a dashed line.}
\end{figure}
\begin{figure}
\centerline{
\psfig{figure=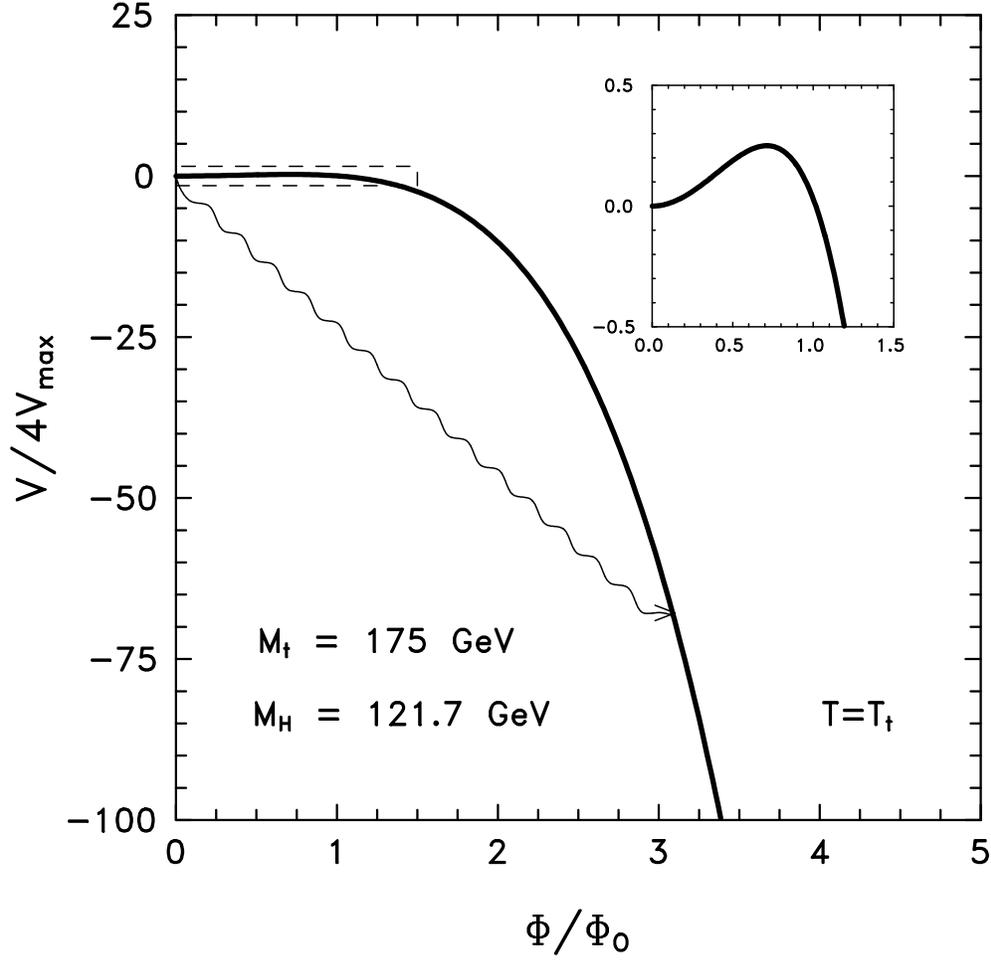,height=13cm,bbllx=6.cm,bblly=.cm,bburx=15.5cm,bbury=13cm}}
\caption{Plot of the effective potential at $T_t=2.5\times 10^{15}$ GeV,
for the same values of $M_t$ and $M_H$ as in Fig.~1, normalized with respect
to its maximum value, as a function of $\phi$, arbitrarily normalized with
$\phi_0=6.0\times 10^{15}$ GeV. The arrow indicates the value of the bounce
solution $\phi_B(0)$.}
\end{figure}
\begin{figure}
\centerline{
\psfig{figure=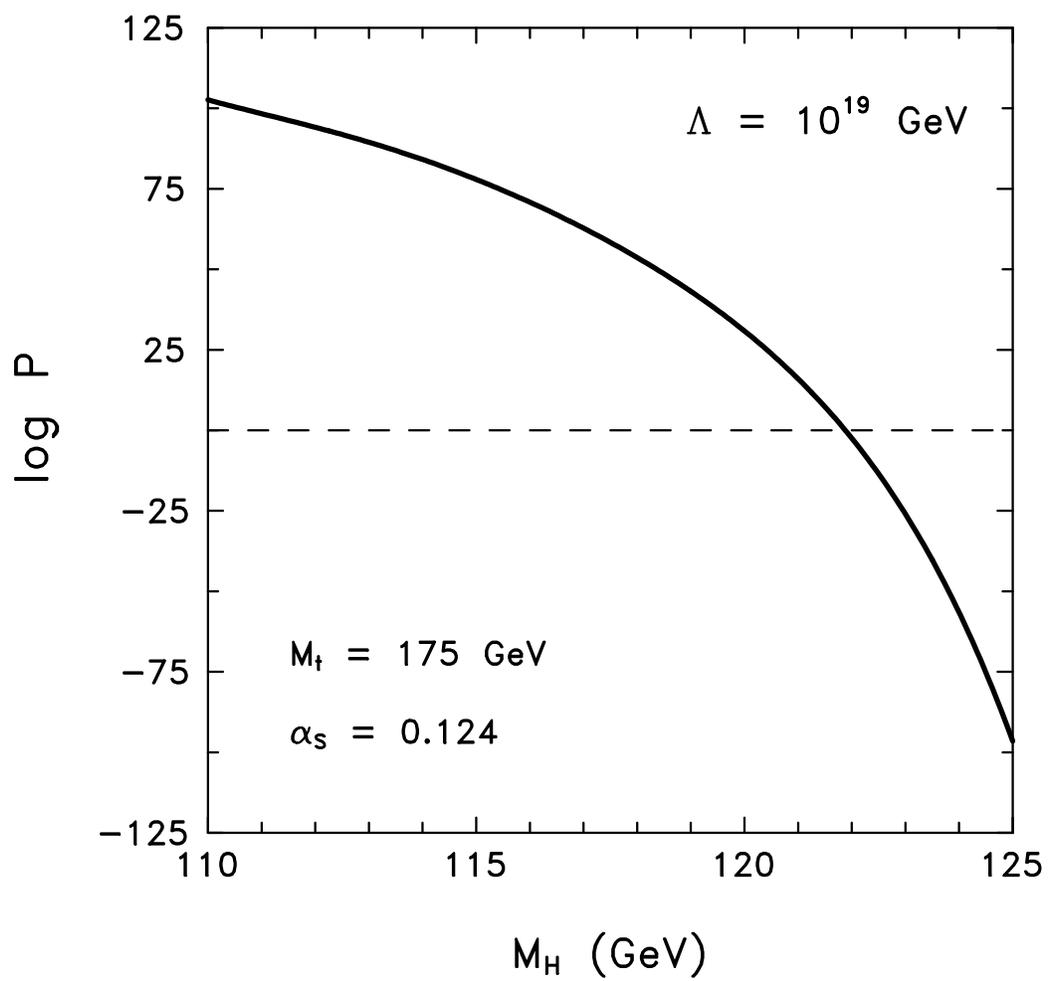,height=13cm,bbllx=6.cm,bblly=.cm,bburx=15.5cm,bbury=13cm}}
\caption{Plot of the logarithm of the total probability ($\log P$)
as a function of $M_H$ for $M_t=175$ GeV.}
\end{figure}
\begin{figure}
\centerline{
\psfig{figure=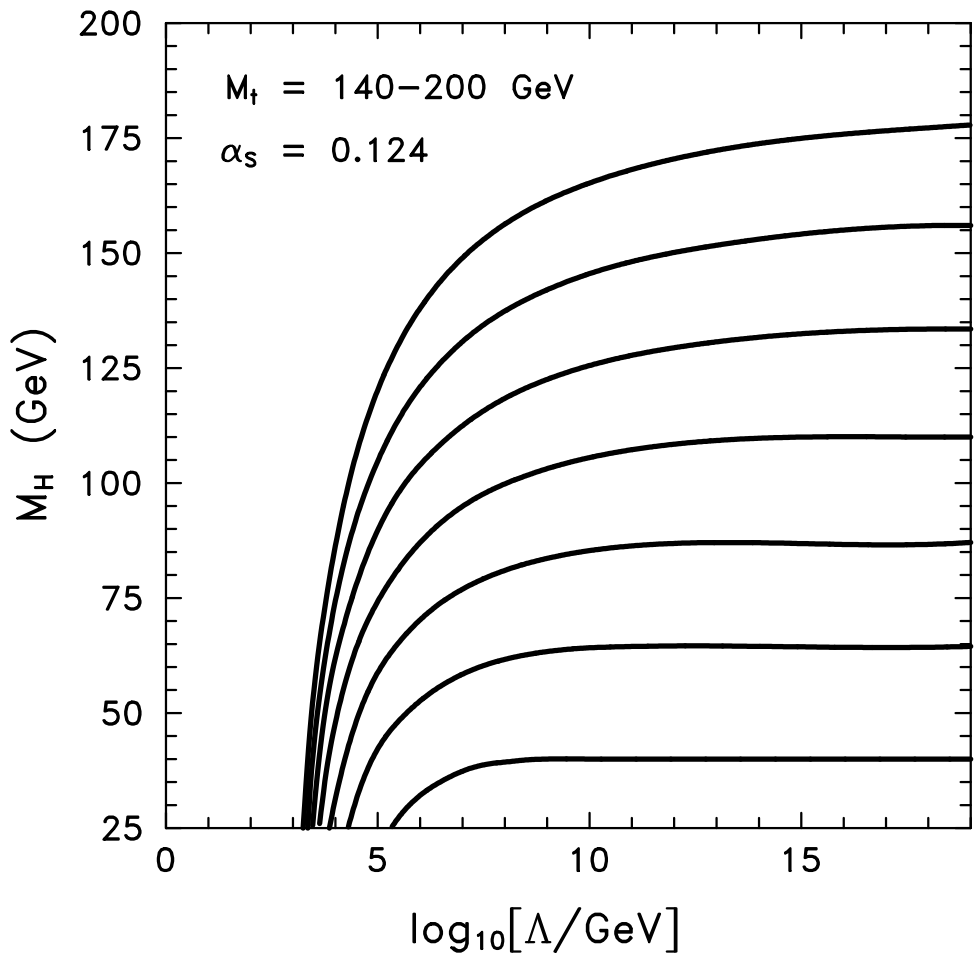,height=13cm,bbllx=6.cm,bblly=.cm,bburx=15.5cm,bbury=13cm}}
\caption{Lower bounds on $M_H$ as a function of the SM cutoff
$\log_{10}[\Lambda/{\rm GeV}]$ for $\alpha_S(M_Z)=0.124$ and $M_t$ from 140
GeV (lower curve) to 200 GeV (upper curve), step~=~10 GeV.}
\end{figure}
\begin{figure}
\centerline{
\psfig{figure=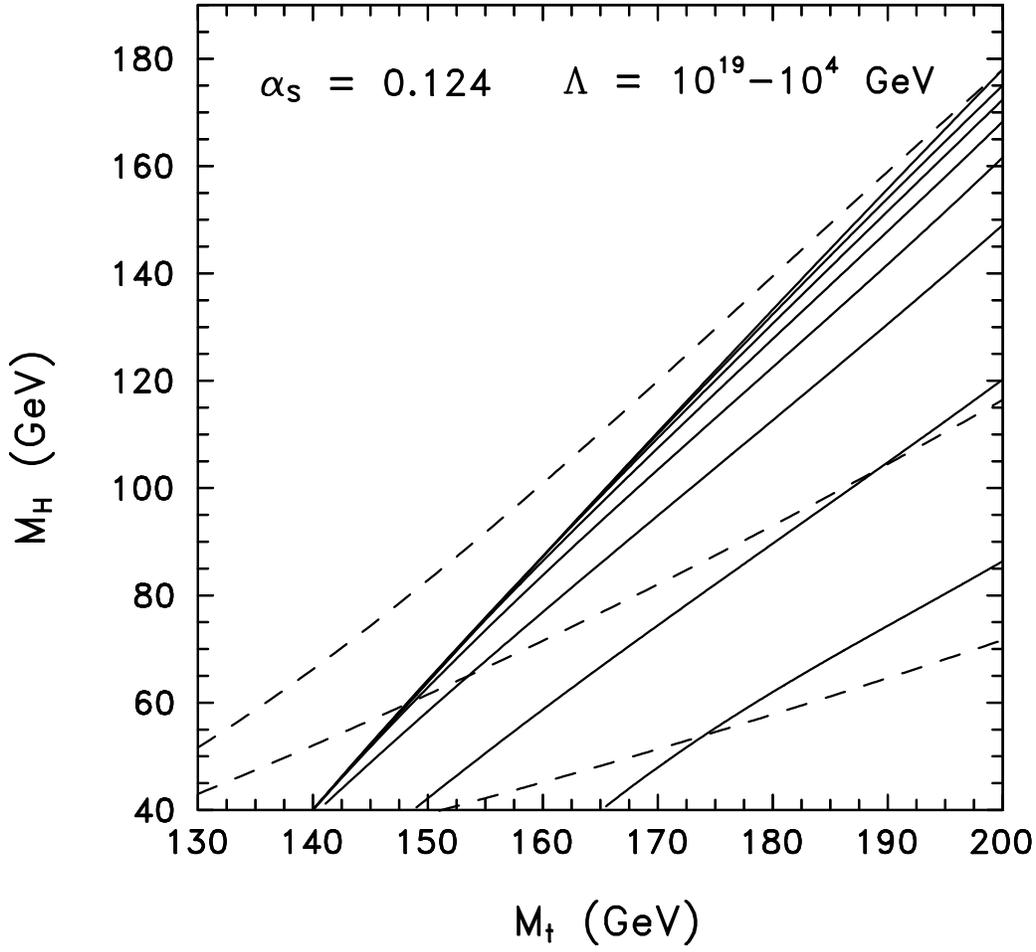,height=13cm,bbllx=6.cm,bblly=.cm,bburx=15.5cm,bbury=13cm}}
\caption{Lower bounds on $M_H$ as a function of $M_t$ for
$\alpha_S(M_Z)=0.124$ and $\Lambda=10^4$~GeV (lower solid line), $10^5$ GeV,
$10^7$ GeV, $10^9$ GeV, $10^{11}$ GeV, $10^{13}$ GeV, $10^{15}$ GeV
and $10^{19}$ GeV (upper solid line). The dashed lines are
the absolute stability bounds for $\Lambda=10^3$ GeV (lower dashed line),
$10^4$ GeV and $10^{19}$ GeV (upper dashed line).}
\end{figure}
\begin{figure}
\centerline{
\psfig{figure=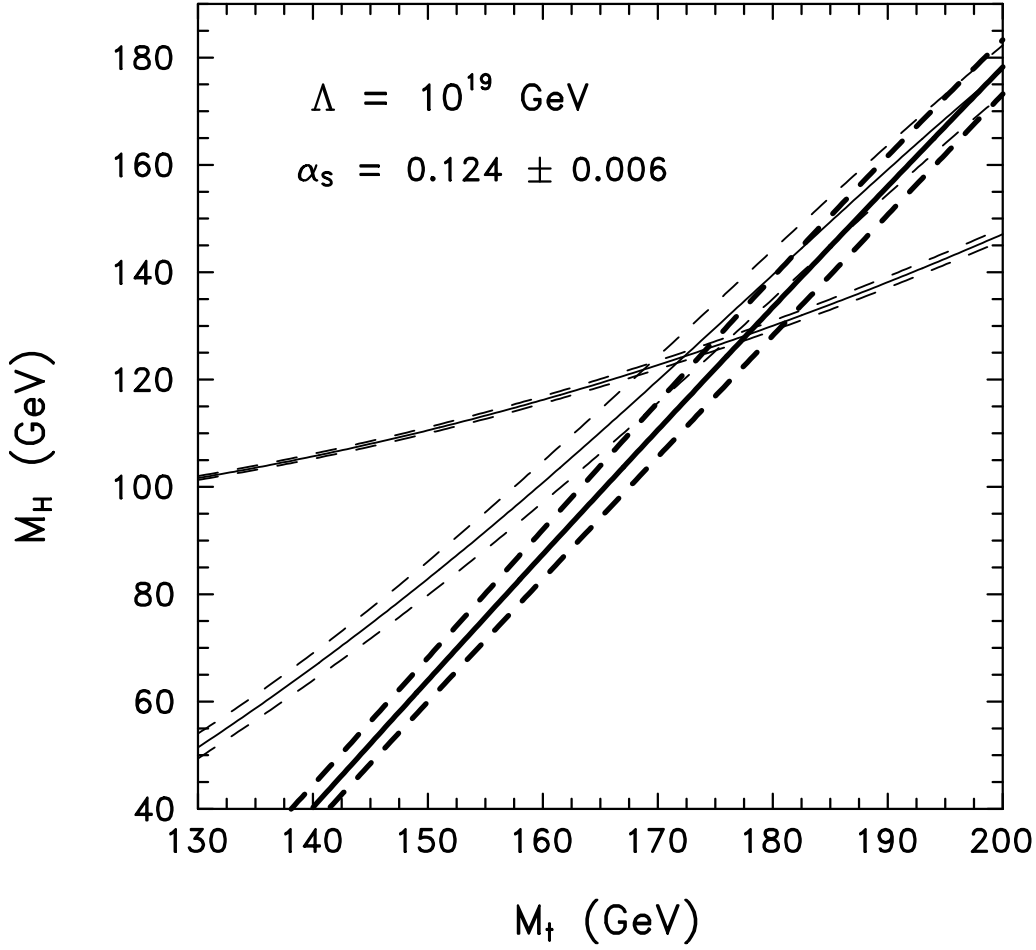,height=13cm,bbllx=6.cm,bblly=.cm,bburx=15.5cm,bbury=13cm}}
\caption{Diagonal lines: SM lower bound on $M_H$ (thick lines) as a function
of $M_t$ for $\Lambda=10^{19}$ GeV and $\alpha_S=0.124$ (solid),
$\alpha_S=0.118$ (upper dashed), $\alpha_S=0.130$ (lower dashed). The
corresponding bounds for the absolute stability requirement are the
diagonal thin lines. Transverse (thin) lines: MSSM upper bounds on $M_H$
for $\Lambda_{\rm SUSY}=1$ TeV and $\alpha_S$ as in the diagonal lines.}
\end{figure}
\begin{figure}
\centerline{
\psfig{figure=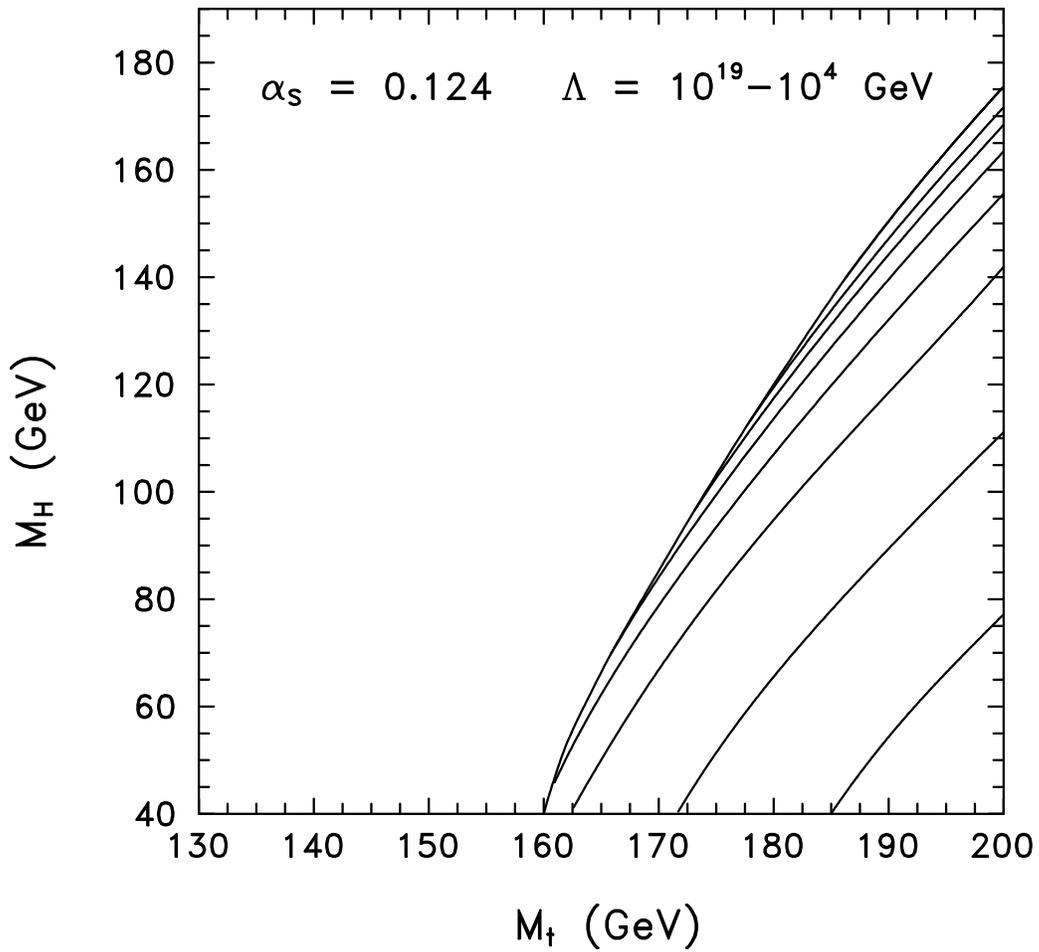,height=13cm,bbllx=7.5cm,bblly=.cm,bburx=17.cm,bbury=13cm}}
\caption{Lower bounds on $M_H$ as a function of $M_t$, for
the values of $\alpha_S(M_Z)$ and $\Lambda$ shown as
solid lines in Fig.~7,
from the requirement of slow quantum tunnelling
(compared with the present Universe expansion rate)
from the electroweak minimum at $T=0$.}
\end{figure}

\end{document}